\documentclass[twocolumn, aps,prl, floatfix, superscriptaddress]{revtex4-1}
\usepackage{graphicx}

\usepackage{color}\usepackage{enumerate}
\usepackage{amsmath}\usepackage{amssymb}\usepackage{stmaryrd}
\usepackage{multirow}
\usepackage{hhline}
\usepackage{tikz}
\usetikzlibrary{matrix}

\usepackage{tcolorbox, accents}
\usepackage{tikz-cd}
\usetikzlibrary{cd}
\usepackage{longtable,booktabs}
\usepackage[colorlinks=true,citecolor=cyan,linkcolor=magenta]{hyperref}
\usepackage{url}
\usepackage{color}\usepackage{enumerate} 
\usepackage{amsmath}\usepackage{amssymb}\usepackage{stmaryrd}
\usepackage{multirow}
\usepackage{float}
\usepackage{mathrsfs}
\usepackage{bbold}
\usepackage[toc,page]{appendix}
\usepackage{longtable,booktabs}
 \usepackage{epsfig}
\usepackage{graphicx} 
\usepackage{feynmf}
\usepackage{psfrag} 
\usepackage{slashed}
\usepackage{mathtools}
\usepackage{amsmath,amssymb} 
\usepackage{colordvi} 
\usepackage{hyperref}
\usepackage{setspace}
\usepackage[all]{hypcap}
\usepackage{natbib}
\usepackage{subfigure}

\definecolor{dcyan}{rgb}{0.0, 0.55, 0.55}
\definecolor{brred}{rgb}{0.8, 0.25, 0.33}
\hypersetup{
   bookmarks=true,         
   unicode=false,          
 pdftoolbar=true,        
    pdfmenubar=true,        
    pdffitwindow=false,     
    pdfstartview={FitH},    
    pdftitle={My title},    
    pdfauthor={Author},     
    pdfsubject={Subject},   
    pdfcreator={Creator},   
    pdfproducer={Producer}, 
    pdfkeywords={keyword1} {key2} {key3}, 
    pdfnewwindow=true,      
    colorlinks=true,       
    linkcolor=brred,          
    citecolor=dcyan,        
    filecolor=magenta,      
    urlcolor=brred           
}

\newcommand{\pd}{\phantom{\dagger}}

\usepackage{natbib} 

\begin{document}
\title{Non-Hermitian topology in monitored quantum circuits}
\author{Christoph Fleckenstein}
\affiliation{Department of Physics, KTH Royal Institute of Technology, Stockholm, 106 91 Sweden}

\author{Alberto Zorzato}
\affiliation{Department of Physics, KTH Royal Institute of Technology, Stockholm, 106 91 Sweden}

\author{Daniel Varjas}
\affiliation{Department of Physics, Stockholm University, AlbaNova University Center, 106 91 Stockholm, Sweden}

\author{Emil J. Bergholtz}
\affiliation{Department of Physics, Stockholm University, AlbaNova University Center, 106 91 Stockholm, Sweden}

\author{Jens H. Bardarson}
\affiliation{Department of Physics, KTH Royal Institute of Technology, Stockholm, 106 91 Sweden}

\author{Apoorv Tiwari}
\affiliation{Department of Physics, KTH Royal Institute of Technology, Stockholm, 106 91 Sweden}

\begin{abstract}
We demonstrate that genuinely non-Hermitian topological phases and corresponding topological phase transitions can be naturally realized in monitored quantum circuits, exemplified by the paradigmatic non-Hermitian Su-Schrieffer-Heeger model. 
We emulate this model by a 1D chain of spinless electrons evolving under unitary dynamics and subject to periodic measurements that are stochastically invoked. 
The non-Hermitian topology is visible in topological invariants adapted to the context of monitored circuits.
For instance, the topological phase diagram of the monitored realization of the non-Hermitian Su-Schrieffer-Heeger model is obtained from the biorthogonal polarization computed from an effective Hamiltonian of the monitored system.
Importantly, our monitored circuit realization allows direct access to steady state biorthogonal expectation values of generic observables, and hence, to measure physical properties of a genuine non-Hermitian model. 
We expect our results to be applicable more generally to a wide range of models that host non-Hermitian topological phases.
\end{abstract}

\maketitle

\emph{Introduction.}---The interplay between non-integrable unitary quantum dynamics, which evolve generic states into volume law entangled states, and measurements, which effectively reduce quantum entanglement, have recently been shown to facilitate novel `measurement-induced phase transitions'   \cite{Nahum_2019, Fisher_2018, Smith_2019, Fisher_2019, Huse_2020, Huse_2020B, Ludwig_2019, Altman_2020, Schomerus_2019, Zhu_2020, Ludwig_2020, DeLuca_2019, Altman_2020B, Qi_2020, Pixley_2020, Chen_2020, Yuto_2020, Peter_2020, Barkeshli_2021, Timothy_2021, Lucas_2020, Ruhman_2021, Pal_2020, Schomerus_2020, lavasani2021topological}. 
A unitary quantum circuit interspersed with local measurements, each invoked with a classical probability $p$, undergoes a measurement-induced phase transition  visible in the entanglement dynamics of typical product states evolved by the quantum circuit.
Various aspects, such as emergent conformal symmetry \cite{Fisher_2019} and effective statistical mechanics descriptions \cite{Ludwig_2019, Ludwig_2020, li2021statistical}, of the  transition as a function of $p$ have already been understood. 
In this work, we show that in addition to entanglement transitions, monitored circuits can realize topological phase transitions, in particular between non-Hermitian topological phases. 
Such topological transitions in monitored circuits are visible in topological diagnostics such as those related to the topological bulk-boundary correspondence.

Non-Hermitian topological systems have recieved considerable attention \cite{Ueda_2018,Fu_2018, Sato_2019,Bergholtz_2018, Zhong_2018, Bergholtz_2021} in the recent past. 
In an important development, the landmark Altland-Zirnbauer classification \cite{AZ_1997,Kitaev_2009, Ryu_2008, Ryu_2010}, which provides an exhaustive topological classification of free fermion and Bogoliubov-de-Gennes systems with combinations of particle-hole, time-reversal and chiral symmetry, has been extended to non-Hermitian systems \cite{Sato_2019}.
Additionally, the bulk-boundary correspondence and bulk topology for candidate models of several topological non-Hermitian phases have been explored \cite{Bergholtz_2018, Vatsal_2019, Okuma_2020, Zhong_2018, Zhong_2018B, Lieu2018, Regnault_2019, Bergholtz_2019, Bergholtz_2021}.
%
These models exhibit various phenomena, such as localization transitions, non-Hermitian skin effect and the emergence of exceptional points, unique to non-Hermitian systems.
%
Among these models, the non-Hermitian Su-Schrieffer-Heeger (nH-SSH) model remains a paradigmatic example that illustrates various novel features of non-Hermitian topology \cite{SSH_original1, SSH_original2, Lee_2016, Lieu2018, Zhong_2018B, Bergholtz_2019, Shuichi_2019, Okuma_2020}. 
\begin{figure}[t]
    \centering
    \includegraphics[scale=0.43]{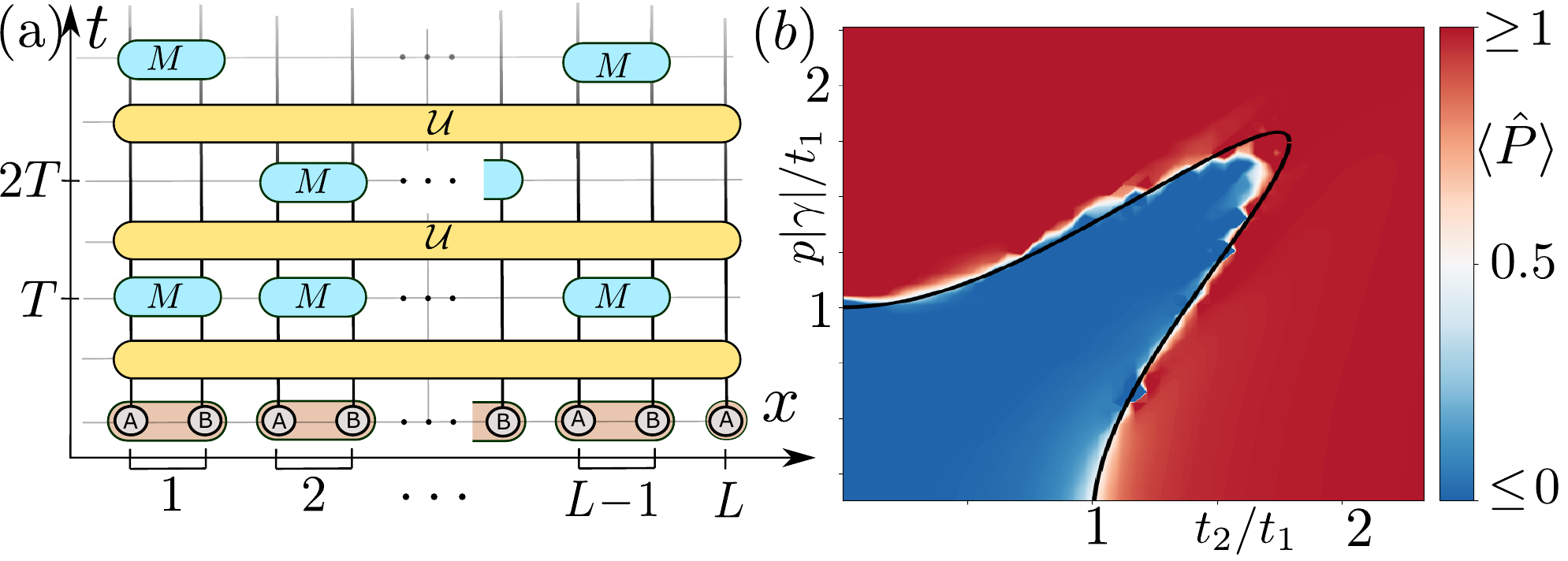}
    \caption{(a) A schematic of the applied sequence of unitary time-evolution ($\mathcal U$) and measurement ($M$) for a chain with a single $A$ atom on the right end, and (b) topological phase diagram of $\mathcal{H}_{\mathrm{eff}}$ [defined in Eq.~\eqref{eq:Heff_from_circuit})] associated with a circuit of depth $N=200$ cycles;
    we display the biorthogonal polarization $\hat{P}$, defined in Eq.~\eqref{eq:polarization}, of the mid-gap state as a function of $p|\gamma|/t_1$
    and $t_2/t_1$, where $p$ is the classical probability of measurement. The solid black line indicates the theoretically expected non-Hermitian topological phase transition at $t_1^4=(t_2^2-\vert p\gamma \vert^2)^2+(2p\gamma_It_2)^2$ . We attribute the small deviations to statistical fluctuations $\sim \mathcal{O}(T\gamma /\sqrt{N})$ emerging from the probabilistic circuit. Further parameters are $L=26$, $Tt_1 =0.001$, and $\gamma/t_1=2.5+0.4i$.
    }
    \label{fig:schematic}
\end{figure}

Alongside theoretical progress there have been several developments in realizing non-Hermitian physics in experimental platforms.
These include photonic \cite{Schomerus_2013,Hodaei_2014, Longhi_2015, Zeuner_2015,Simon_2015, Weimann_2017y, Amo_2017, Hodaei_2017, Cerjan_2019}, phononic (mechanical) \cite{Shi_2016, Brandenbourger_2019, Hatsugai_2019, Ghatak2020} and topoelectric circuit \cite{Neupert_2020} setups, where the evolution of the electric, acoustic and current modes determined by Maxwell's, Newton's and the generalized Ohm's equation, respectively, are formally equivalent to the Schr\"{o}dinger equation and can furthermore be made non-Hermitian owing to the ubiquity of dissipation, loss and gain in such systems. 
Despite these experimental milestones, examples of truly quantum physical realizations of non-Hermitian phases remain scarce.
Additionally a quantum-mechanical protocol to simulate the biorthogonal expectation value of generic operators---standard in the description of non-Hermitian systems \cite{Brody_2013}
%
---is lacking.

Two main avenues expected to generate effective non-Hermiticity in the single particle evolution of a quantum particle are (i) dissipative processes involving interactions with an external bath \cite{Diehl_2011, Bardyn_2013, Lee_2016, Duan_2017, Diehl_2018, Cooper_2020, Nori_2020, Diehl_2021} and (ii) strong interactions that lead to finite lifetimes of quasiparticles appearing as a non-Hermitian correction to the non-interacting Hamiltonian via the self energy \cite{kozii2017nonhermitian, Torres_2018, Fu_2018, Kawakami_2018,PhysRevResearch.1.012003, Hatsugai_2020, Fu_2019, Yoshida_2021, Fu_2020}. 
In the context of dissipative dynamics, it has been shown \cite{Nori_2020,Okuma_2021} that post-selected quantum trajectories can be modelled as non-unitary dynamics governed by an effective non-Hermitian Hamiltonian.
%
In such a setup the quantum trajectory without any quantum jumps is post-selected. 

In this work, we propose an alternate route towards realizing non-Hermitian topological phases in monitored quantum circuits. 
Instead of post-selecting a quantum trajectory without quantum jumps, we post-select trajectories with definite measurement outcomes for generalized measurements
\cite{Nahum_2019, Fisher_2018, Smith_2019, Fisher_2019, Huse_2020B, Ludwig_2019, Altman_2020, Schomerus_2019, Zhu_2020, Ludwig_2020, DeLuca_2019, Altman_2020B, Qi_2020, Pixley_2020, Chen_2020, Yuto_2020, Peter_2020, Barkeshli_2021, Pal_2020, Schomerus_2020, lavasani2021topological}.
%
The measurements are implemented locally with some classical probability $p$ and punctuate otherwise unitary dynamics (cf.~Fig.~\ref{fig:schematic}(a)).
Using the nH-SSH model as a case study, we show that such quantum circuits can be interpreted as generators of non-Hermitian time-evolution. As in (Hermitian) Floquet topological systems \cite{Oka2009,Kitagawa2010,Lindner2011,Liu2013,Yan2016,Loss2017,Fleckenstein2020}, the corresponding \textit{effective} Hamiltonian can be attributed to distinct topological properties and phases.
%
In our system, the Hermitian part of a given non-Hermitian topological Hamiltonian is incorporated in the unitary component of the monitored circuit while the non-Hermiticity is modelled using generalized measurements. 
The corresponding topological phase diagram can be extracted directly from a non-Hermitian effective Hamiltonian using the biorthogonal bulk-boundary correspondence as a diagnostic. 
In fact, these biorthogonal expectation values can be accessed directly in the monitored circuit via auto-correlators. 
This offers a novel way to obtain physically relevant quantities of genuine non-Hermitian models.

\emph{Non-Hermitian Su-Schrieffer-Heeger model.}---The nH-SSH model describes noninteracting spinless electrons hopping on a 1D lattice.
Each unit-cell contains two orbitals denoted $A$ and $B$.
The Hamiltonian $\mathcal H(t_1,t_2,\gamma)=\mathcal H_{1}(t_1)+ \mathcal H_{2}(t_2,\gamma)$ includes intra and inter-cell hopping terms 
\begin{align}
 \mathcal H_1(t_1)=&\; \sum_{j} t_1 \left[ c^{\dagger}_{j,B}c^{\pd}_{j+1,A}+  c^{\dagger}_{j+1,A}c^{\pd}_{j,B} \right]. \\
\mathcal H_2(t_2,\gamma)=&\; \sum_{j}\left[ (t_2 \!+\!\gamma) c^{\dagger}_{j,A}c^{\pd}_{j,B}+ (t_2\! -\!\gamma^{*}) c^{\dagger}_{j,B}c^{\pd}_{j,A}\right]\nonumber 
\end{align}
where $c_{j,A}^{\dagger}$ ($c_{j,B}$) are fermionic creation (annihilation) operators on the $A$ ($B$) atom of site $j$. $\gamma=\gamma_{R}+i\gamma_I$ and the parameters $t_{1,2}$ and $\gamma_{I,R}$ are real-valued such that $\mathcal H_{1}$ is a Hermitian operator while $\mathcal{H}_{2}$ has both Hermitian and anti-Hermitian terms proportional to $t_2$ and $\gamma$ respectively.
%

%
%
Complex $\gamma$ leaves the system with only chiral symmetry and reduces the number of topologically distinct phases from three (at $\gamma \in \mathbb{R}$) to two, bounded by $t_1^4=(t_2^2-\vert \gamma \vert^2)^2+(2\gamma_It_2)^2$ (cf.\ black solid line in Fig.~\ref{fig:schematic}(b)).
 %
 %
 %
 The spectrum of the Hamiltonian depends sensitively on the choice of boundary conditions \cite{Bergholtz_2018,Zhong_2018, Okuma_2020}. 
 The open chain has a bulk line-gap and potentially a single mid-gap zero mode state depending on the boundary termination, while the periodic chain is either gapped or has exceptional points. 
 For concreteness, we focus on an open 1D chain with $L$ unit-cells which terminates on an $A$-orbital at both ends. 
 For this setup, one always obtains a single zero-mode localized on either end of the chain defining two distinct topological phases,  distinguished by the biorthogonal polarization \cite{Bergholtz_2018} 
 $ \mathcal P=  \langle \Psi_{L}| \hat{P} |\Psi_{R}\rangle $ where $|\Psi_{L(R)}\rangle$ are the left (right) biorthogonal zero-mode eigenstates and
\begin{align}
\label{eq:polarization}
    \hat{P}=&\;  \frac{1}{L}\sum_{j=1}^{L}j\Pi_j ,
\end{align}
where $   \Pi_{j}=c^{\dagger}_{j,A}|0\rangle \langle 0| c_{j,A}+c^{\dagger}_{j,B}|0\rangle \langle 0| c_{j,B}$
is the projection operator on the $j^{\text{th}}$ unit-cell. 
A topological phase transition is characterized by a jump in $\mathcal P$ \cite{Bergholtz_2018} and is equivalent to a change in the winding number defined in a generalized Brillouin zone \cite{Zhong_2018,Vatsal_2019}.

\emph{Monitored circuit and measurement protocol.}---In what follows, we realize the nH-SSH model as a quantum circuit constructed from two ingredients: (i) a unitary evolution operator that corresponds to the Hermitian part of the nH-SSH model and (ii) measurement operators executing generalized measurements that act locally on the unit-cells of the nH-SSH chain and emulate the non-Hermiticity of the model, cf.\ Fig.~\ref{fig:schematic}(a). 
The unitary time-evolution operator is obtained by simply setting $\gamma=0$ in the nH-SSH evolution operator $\mathcal U(T)=\exp\left\{-iT\mathcal H(t_1,t_2,0)\right\}$ (we set $\hbar = 1$).
A measurement on the $j^{\text{th}}$-unit-cell of the 1D chain is defined by an operator $M_{j,+}$, which is a member of a set of Hermitian Kraus operators $\left\{M_{j,\alpha}\right\}$ with $\alpha\in \left\{0,\pm\right\}$.
The Kraus operators define a completely-positive trace-preserving map that maps $\rho$, the density matrix of the fermionic chain, to
$\sum_{\alpha}M_{j,\alpha}\rho M_{j,\alpha}^{\dagger}$. 
Explicitly, these take the form
\begin{align}
 M_{j,\pm}=&\;\sqrt{\frac{1}{2(1+|\gamma T|^2)}}\left[1\pm iT\left(\gamma^{*} c^{\dagger}_{j,B}c^{\pd}_{j,A}
 -\gamma 
 c^{\dagger}_{j,A}c^{\pd}_{j,B}
 \right)\right], \nonumber \\
 M_{j,0}=&\;\sqrt{\frac{|\gamma T|^2}{1+|\gamma T|^2}}(1-c_{j,A}^{\dagger}c_{j,A}-c_{j,B}^{\dagger}c_{j,B}),
\end{align}
and satisfy the trace-preserving condition of a quantum map $\sum_{\alpha}M_{j,\alpha}^{\dagger}M_{j,\alpha}^{\pd}=1$. 
%
%
Note that for $|\gamma T|=1$, $M_{j,\pm}$ and $M_{j,0}$ implement projective measurements in the single particle sector.
We consider a measurement protocol that post-selects the `$+$' outcome, i.e., the measurement maps the state $|\Psi\rangle$ to $M_{j,+}|\Psi\rangle /||M_{j,+}|\Psi\rangle||$.
%
Such a measurement outcome occurs with probability $||M_{j,+}|\Psi\rangle||$ and in the limit $|\gamma T| \to 0$ (the regime we work in) only the measurement outcomes corresponding to $M_{j,\pm}$ survive.
Post-selecting the $M_{j,+}$ outcome over $n$ measurements therefore involves performing $~2^{n}$ successive experiments.
Such a generalized measurement protocol can be implemented by coupling the $j^{\text{th}}$-unit-cell in the 1D chain to a ``macroscopic" ancilla qubit  $|\sigma\rangle_j$ as follows: 
First, the total system is prepared in a tensor product state $|\Psi\rangle \otimes |0\rangle_j$, where $|0\rangle_j$ is a reference qubit state and $|\Psi\rangle$ is the initial state of the 1D chain.
Then, one acts with the unitary $\mathcal U_{M_j}:|\Psi\rangle \otimes |0\rangle_j \mapsto \sum_{\alpha=\pm}M_{j,\alpha}|\Psi\rangle \otimes |\alpha \rangle_j$ \footnote{The operator $\mathcal U_{M_j}$ is unitary up to an error of $\mathcal O(|\gamma T|^2)$, i.e. $\mathcal U_{M_j}^{\dagger}\mathcal U_{M_j}^{\pd}=1-\mathcal O(|\gamma T|^2)$.},
where $|\pm\rangle_j$ denote the $\sigma_z$ eigenstates of the qubit. 
Finally, a projective measurement for the ancilla qubit onto $|+\rangle_j$ is carried out.
The normalized post-measurement state thus obtained is precisely $M_{j,+}|\Psi\rangle /||M_{j,+}|\Psi\rangle||$.
Unitary evolution for a single time step $T$ followed by a simultaneous post-selective measurement on all the unit-cells is described by the operator $F=   (\prod_{j}M_{j,+}) \cdot  \mathcal U(T).$
%
Instead of carrying out the measurement on every unit-cell, we consider a protocol wherein the measurement operation is invoked on each unit-cell with a classical probability $p\in [0,1]$. 
Thus one obtains a probabilistic quantum circuit operator that acts on a state $|\Psi\rangle$ as
\begin{align}
    F(p, \boldsymbol{s})|\Psi\rangle=\frac{
    \Big(\prod_{j}M_{j,+}^{s_j}\Big)\cdot \mathcal U(T) |\Psi\rangle}{\Big|\Big|\Big(\prod_{j}M_{j,+}^{s_j}
    \Big)
    \cdot \mathcal U(T) |\Psi\rangle\Big|\Big|},
\end{align}
 with $s_j=0$ or $1$ with probability $1-p$ and $p$ respectively. 
Correspondingly, a depth $N$ monitored circuit with measurement probability equal to $p$ is defined as 
\begin{align}
    F_{N}\left(p,\left\{\boldsymbol{s}\right\}\right):=F(p,\boldsymbol{s}_N)\cdot F(p,\boldsymbol{s}_{N-1})\cdots F(p,\boldsymbol{s}_1).
\end{align}
%
\emph{Effective Hamiltonian and Non-Hermitian topology.}---Next, we extract the nH-SSH topological phase diagram directly from the monitored quantum circuit. 
For the deterministic case of $p=1$, the circuit is periodic with period $T$ and we can apply a Floquet-Magnus expansion to obtain a Floquet Hamiltonian $\mathcal H_F$, which governs the dynamics over a unit time-step $T$ $F(p=1,\boldsymbol{s})$\footnote{For $p=1$, $\boldsymbol{s}$ actually becomes superfluous.}. 
In the high-frequency limit, where $T\nu \ll 1$ ($\nu \in \{t_1,t_2,\gamma\}$), the measurement operator $\prod_j M_{j,+}\propto \exp\left\{-iT\mathcal H_2(0,\gamma)\right\}$ and consequently the Floquet Hamiltonian (to leading order in inverse-frequency)
equals the single-particle nH-SSH Hamiltonian, i.e., $\mathcal H_{\text{F}}=\mathcal H(t_1,t_2,\gamma)+ \mathcal{O}(T\nu)$. 
Moreover, $\mathcal H_{\text{F}}$ captures the dynamics of the system for (at least) exponentially long times $\sim \mathrm{exp}[1/(\nu T)]$ \cite{abanin_15,mori_15} \footnote{Since $\mathcal H_F$ is single-particle at high-frequencies, the  notion of thermalization is not well defined. Then $\mathcal{H}_F$ is expected to exceed the exponential lower life-time bound $\exp[1/(\nu T)]$, derived for many-body Hamiltonians}. 
For $p\neq 1$ periodicity is lost; however, an $N$-cycle \textit{effective} Hamiltonian can be defined as 
\begin{align}
    \mathcal{H}_{\text{eff}}^{(N)}(p,\left\{\boldsymbol{s}\right\})=\frac{i}{NT}\sum_{k=1}^{N}\log(F(p,\boldsymbol{s}_{k})).
   \label{eq:Heff_from_circuit}
\end{align}
\begin{figure}
    \centering
    \includegraphics[scale=0.38]{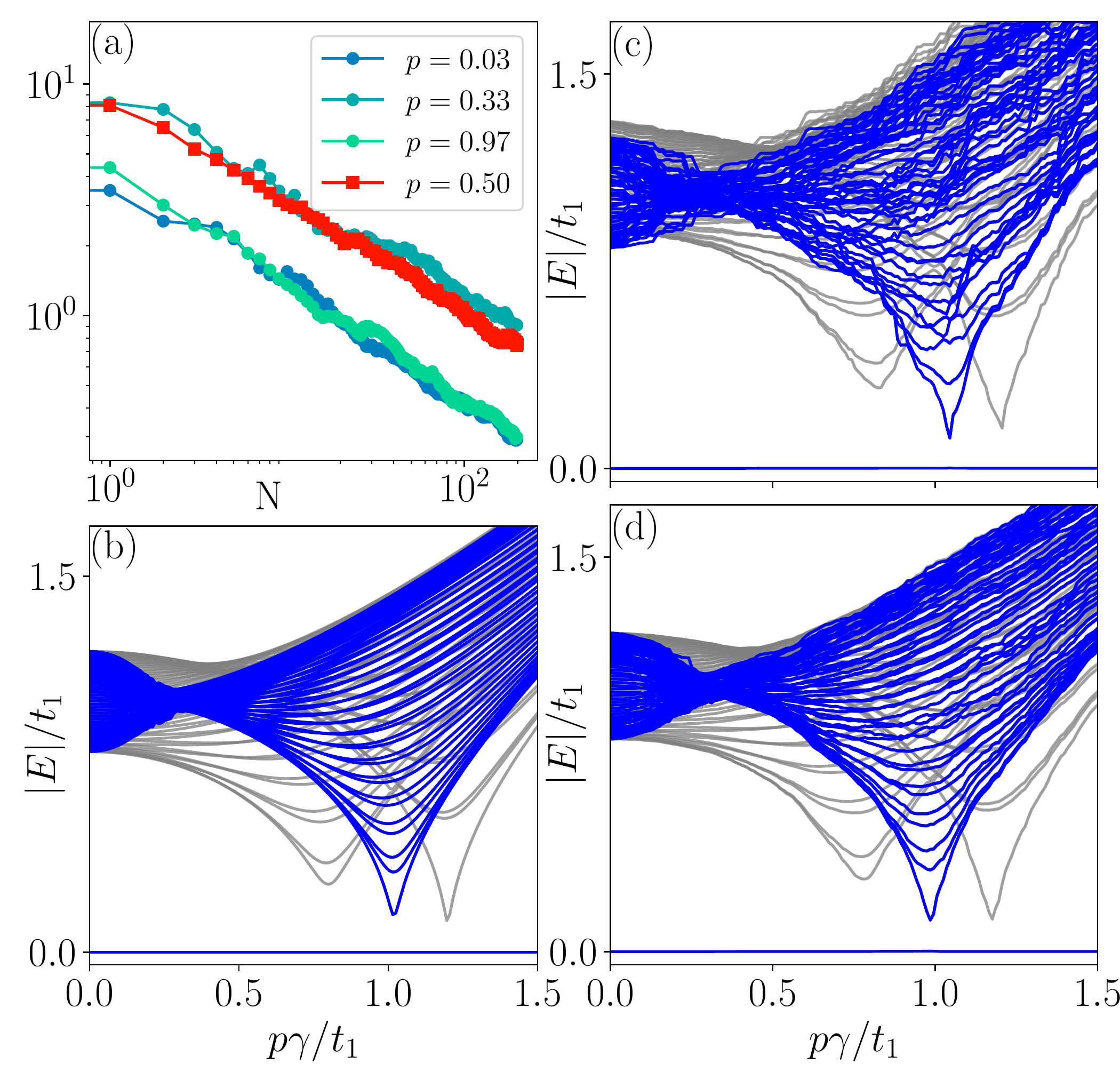}
    \caption{(a) Matrix norm difference $\vert\vert \mathcal{H}_{\mathrm{eff}}^{(N+1)}(p,\{\boldsymbol{s}\}) - \mathcal{H}_{\mathrm{eff}}^{(N)}(p,\bar{\boldsymbol{s}}\})\vert \vert /t_1 $ with two unrelated realizations $\{\boldsymbol{s}\}$ and $\{\bar{\boldsymbol{s}}\}$ computed according to Eq.~\eqref{eq:Heff_from_circuit} for different values of $p$ (green, turquoise and blue circles) and matrix norm difference $\vert\vert \mathcal{H}_{\mathrm{eff}}^{(N)}(p,\{\boldsymbol{s}\})\big\vert_{\gamma} - \mathcal{H}_{\mathrm{eff}}^{(N)}(1,\{\boldsymbol{s}\})\big\vert_{p\gamma} \vert \vert /t_1  $ (red squares).
    (b) Spectrum of the nH-SSH model $\mathcal{H}(t_1,t_2,p\gamma)$ for open (blue) and periodic (grey) boundary conditions. We use a system with complete unit cells in the periodic boundary case to restore translational symmetry. (c)-(d) Spectrum of the effective Hamiltonian associated with a probabilistic monitored circuit at $N=20$ (c) and $N=100$ (d) with open (blue) and periodic (grey) boundary conditions. In (c) and (d) we tune the classical probability $p$ and compute the effective Hamiltonian, while in (b) we tune $\gamma_{\mathrm{eff}}=p\gamma$ correspondingly. In all plots we have $t_2/t_1 = 0.2$, $L=50$, $\gamma /t_1 = 2+0.2i$ and $Tt_1 = 0.001$.}
    \label{fig:norm_diff}
\end{figure}
For increasing $N$, this effective Hamiltonian $\mathcal{H}_{\text{eff}}^{(N)}$ is self-averaging, such that deviations fall off as $1/\sqrt{N}$ according to the central limit theorem, cf.\ Fig.~\ref{fig:norm_diff}(a). 
%
In this limit, the many-cycle effective Hamiltonian recovers the nH-SSH Hamiltonian (in leading order Magnus expansion) up to corrections $\mathcal{O}(T\gamma/\sqrt{N})$
\begin{equation}
    \label{eq:H_eff_nHSSH}
    \mathcal H_{\text{eff}}^{(N)}(p,\left\{\boldsymbol{s}\right\}) =\mathcal{H}(t_1,t_2,p\gamma) + \mathcal{O}(T\gamma/\sqrt{N}) + \mathcal{O}(T\nu).
\end{equation}
%
Consequently, the effective Hamiltonian for a probabilistic circuit $F_N(p,\{\boldsymbol s\})$ at large $N$ only depends on $\gamma p$. This allows to replace a probabilistic circuit with given $\gamma$ and $p$ with an effective deterministic circuit with $\gamma_{\mathrm{eff}}=p\gamma$ (cf. Fig.~\ref{fig:norm_diff}(a)).
%
Hence, the dynamics of the probabilistic circuit is equally well captured by $\mathcal{H}(t_1,t_2,p\gamma)$ up to exponentially long times $\sim \exp[1/(\nu T)]$, given $1\ll N < \exp[1/(\nu T)]$. 
Interestingly, the (low-energy) topological properties of $\mathcal H_{\text{eff}}^{(N)}(p,\{\boldsymbol{s}\})$ match well those of $\mathcal H(t_1,t_2,p\gamma)$ even for fairly small values of $N$ (Fig.~\ref{fig:norm_diff}(b) and (c)), while larger values of $N$ increase the coincidence also at higher energies (Fig.~\ref{fig:norm_diff}(b) and (d)).
Subsequently, we work in the large $N$ regime and use $\mathcal H_{\text{eff}}^{(N)}(p,\{\boldsymbol{s}\})=\mathcal H_{\mathrm{eff}}(p)$, $F_N(p,\{\boldsymbol{s}\})=F_N(p)$ for convenience.
Since the circuit $F_{N}(p)$ conserves particle number microscopically with each circuit component individually commuting with the particle number operator, it is convenient to block-diagonalize the circuit and work in the single-particle Hilbert space. 
The topological phase diagram of the circuit can then be obtained from the spectrum of $ \mathcal H_{\text{eff}}(p)$ using the biorthogonal polarization $\mathcal P$ as a topological diagnostic (see Fig.~\ref{fig:schematic}(b)).

\emph{Monitored state evolution and non-Hermitian topology.}---Thus far, we have obtained the topological phase diagram from the effective Hamiltonian defined using the monitored quantum circuit.
A natural question is whether the topology can also be inferred directly from the evolution of states under the monitored quantum circuit?
Since the monitored circuit emulates time-evolution with a genuine non-Hermitian Hamiltonian, exhibiting the non-Hermitian skin effect, all initial states eventually terminate in a steady state corresponding to the eigenstate with the largest imaginary energy eigenvalue and a nonzero overlap with the initial state.
This can be directly deduced from an eigenstate expansion of the effective non-Hermitian evolution operator
\begin{equation}
F_{N}(p):\vert \psi_0\rangle \mapsto \sum_{m}\frac{e^{-iNTE_m} \vert \Psi_{R,m} \rangle \langle\Psi_{L,m} \vert \psi_0\rangle}{||F_{N}(p)\vert\psi_0\rangle||},
\end{equation}
where we have used the biorthogonal decomposition of $ \mathcal H_{\text{eff}}(p)$ 
with complex eigenenergies $\left\{E_{m}\right\}$ ordered according to decreasing imaginary component such that $\text{Im}(E_m)\geq\text{Im}(E_{m+1})$. 
Clearly, in the limit that $NT\text{Im}(E_1-E_2)\gg 1$ and assuming $\langle \Psi_{L,1} \vert \psi_0 \rangle \neq 0$, the initial state $|\psi_0\rangle$ evolves into the steady state $|\Psi_{R,1}\rangle$ under the monitored circuit action, $ F_{N}(p)\vert \psi_0\rangle \approx  \vert \Psi_{R,1} \rangle  \langle\Psi_{L,1} \vert \psi_0\rangle$, provided the evolution time $N\nu T$ is less than the lifetime $\exp[1/(\nu T)]$ of the effective Hamiltonian.
%
\begin{figure}
    \centering
    \includegraphics[scale=0.34]{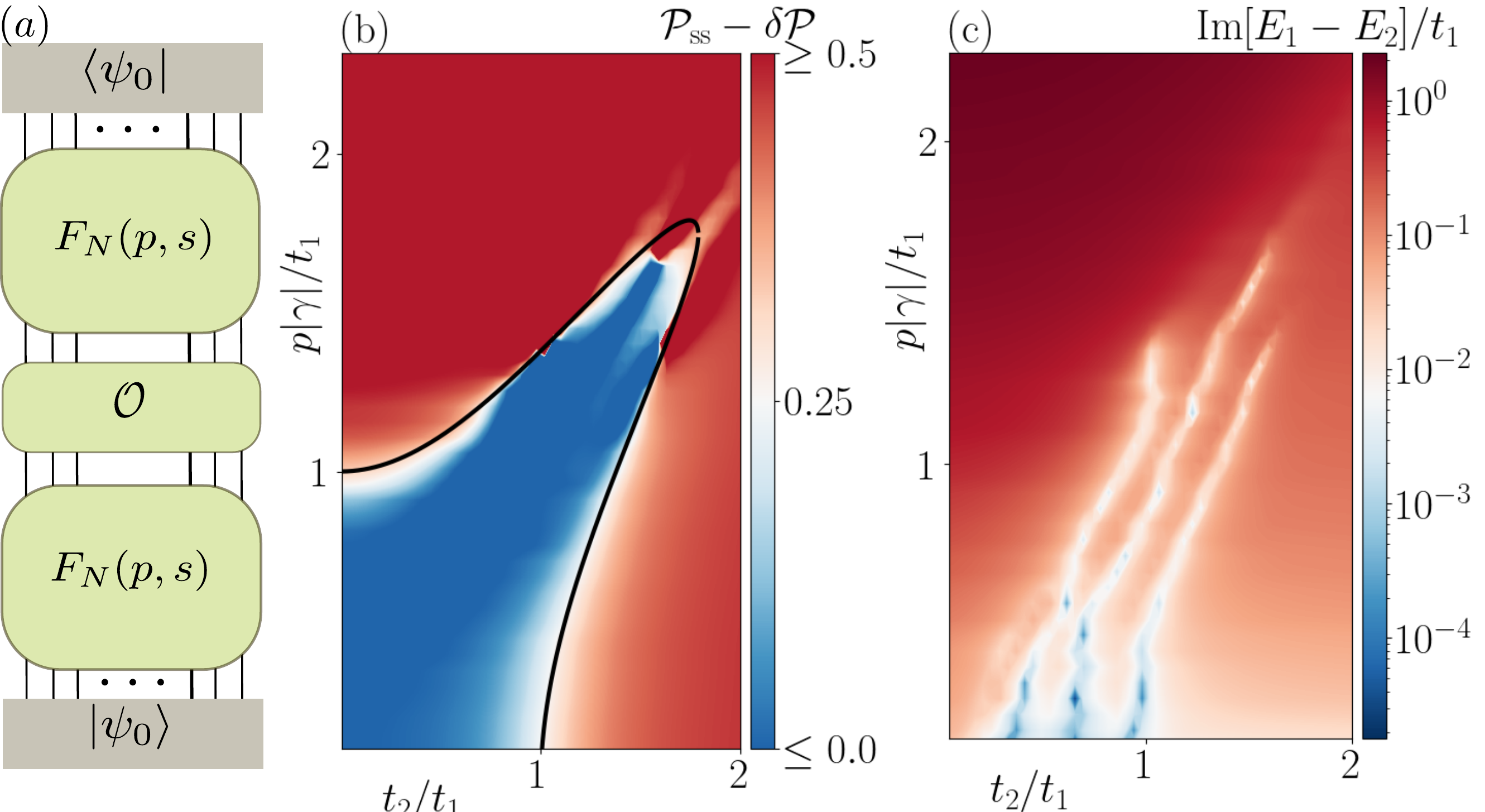}
    \caption{(a) A schematic of a circuit realizing the biorthogonal steady state expectation value of the operator $\mathcal{O}$. (b) Biorthogonal polarization $\mathcal{P}_{\mathrm{ss}}-\delta\mathcal{P}$ of the steady state and (c) corresponding imaginary gap $\mathrm{Im}(E_1-E_2)/t_1$, associated with a monitored circuit of $N=150$ cycles. The solid black line in (b) indicates the analytical result for the non-Hermitian phase transition. Further parameters of the plot are $Tt_1= 0.001$, $L=7$, $\gamma/t_1 = 2.5+0.4i$.}
    \label{fig:p_max}
\end{figure}
%

%

To reveal the topological properties of the monitored circuit, we want to probe the mid-gap state, which in the single-particle sector (except for highly fine-tuned initial states) does not correspond to the steady state. 
Owing to an emergent chiral/sublattice symmetry of $\mathcal H_{\text{eff}}(p)$, this problem can be circumvented at half filling $n= L$, where $n$ is the number of electrons in the chain. 
The steady state at half-filling is a Slater determinant of the eigenstates of $\mathcal H_{\text{eff}}(p)$ with imaginary part of the eigenvalue larger or equal to that of the mid-gap state, and is therefore expected to contain information about the topology of the model.
%
Yet, to extract this inherent information, ordinary expectation values may not suffice: the time-evolution associated with the monitored circuit is subject to the non-Hermitian skin effect, rendering ordinary expectation values ineffective as probability currents are not preserved. 
Instead we require \textit{biorthogonal} expectation values. 
An intriguing way to design such a biorthogonal steady state expectation value is through auto-correlators of the monitored circuit. In particular, let us consider the following auto-correlator
\begin{equation}
\label{eq:auto_corr}
    \mathcal{A}(\hat{P},N)=\frac{ \langle \psi_0 \vert F_N(p)\hat{P} F_N(p)\vert \psi_0 \rangle}{ \langle \psi_0 \vert F_{2N}(p)\vert \psi_0 \rangle},
\end{equation}
associated with the circuit in Fig.~\ref{fig:p_max} (a). The denominator ensures biorthogonal normalization.
In the limit $NT\text{Im}(E_1-E_2)\gg 1$  we readily obtain
\begin{equation}
\mathcal{A}(\hat{P},N) = \langle \!\psi_0 \vert \Psi_{R,1} \!\rangle \langle\! \Psi_{L,1} \vert \psi_0 \! \rangle \mathcal{P}_{\mathrm{ss}},
\end{equation}
where $\mathcal{P}_{\text{ss}}=  \langle \Psi_{L,1} \vert \hat{P} \vert \Psi_{R,1} \rangle/\langle \Psi_{L,1} \vert \Psi_{R,1}\rangle$ is the biorthogonal polarization of the steady state. 
Note that $\gamma \in \mathbb{C}$ ensures $\mathrm{Im}(E_1) \neq \mathrm{Im}(E_2)$ for the whole parameter regime [cf.~Fig.~\ref{fig:p_max} (c)]. For real $\gamma$, $\mathrm{Im}(E_1) \neq \mathrm{Im}(E_2)$ can be ensured only with $\gamma > t_2 $. In this case its still possible to trace one of the two topological transitions with the steady state approach.
%
%
{{Since the many-body groundstate at half-filling is a Slater determinant of a topological zero-mode state and remaining bulk single particle eigenstates, the quantization of $\hat{P}$ gets modified.
Subtracting off a constant background $\delta\mathcal{P}=(n-1)/2$ the polarization jumps by $~0.5$ across a topological phase transition.} } 
%
%
In Fig.~\ref{fig:p_max} (b) we display $\mathcal{P}_{\mathrm{ss}}-\delta\mathcal{P}$ for a circuit of $N=150$ driving cycles at half filling. 
Importantly, $\mathcal{P}_{\mathrm{ss}}$ is able to trace the topology of the system. 
As a guide to the eye we display the same phase boundary lines shown in Fig.~\ref{fig:schematic}(b), corresponding to the analytically expected topological phase transition.
Thus, from measuring the auto-correlator $\mathcal{A}(\hat{P},N)$ of the monitored circuit---provided the system is initialized suitably, yet not fine-tuned---it is possible to obtain direct access to biorthogonal expectation values that carry information about the topological properties that can be attributed to the circuit itself. 
This adds a novel and intriguing meaning to non-Hermitian topological systems and provides a measurable signature of biorthogonal expectation values.

\textit{Conclusion.}---In this Letter we have proposed a quantum circuit of free fermions built from interspersed generalized measurements and unitary time-evolution that effectively implements a non-Hermitian SSH model. This minimal model embraces all genuine non-Hermitian ingredients including the non-Hermitian skin effect, which makes it challenging to obtain physically meaningful observables able to trace the topology of the system. Yet, the quantum circuit is able to provide such a topological signature since the auto-correlator of the form of Eq.~\eqref{eq:auto_corr} resembles a steady state biorthogonal expectation value. 

While preparing the manuscript, we became aware of the recent work \cite{kells2021topological}, which also studies effective non-Hermitian topological phases arising in monitored quantum systems. 
While our physical setup is similar to \cite{kells2021topological}, we study observables of a genuine non-Hermitian topological phase exhibiting the skin-effect.
\emph{Acknowledgments.---} This work received funding from the European Research Council (ERC) under the European Union’s Horizon 2020 research and innovation program (grant agreement No.~101001902),  the Swedish Research Council (VR) through grants number 2018-00313, 2019-04736 and 2020-00214, and the Knut and Alice Wallenberg Foundation (KAW) via the Wallenberg Academy Fellows program (2018.0460) and the project Dynamic Quantum Matter (2019.0068).

\bibliography{refs}
\end{document}